\documentclass[twocolumn,showpacs,preprintnumbers,amsmath,amssymb,superscriptaddress]{revtex4-1}
\usepackage{graphicx}
\usepackage{dcolumn}
\usepackage{bm}
\usepackage{CJK, CJKnumb}
\usepackage{color}
\begin{document}
\begin{CJK*}{GBK}{song}

\title{Topologically protected magnetoresistance by quantum anomalous Hall effect}
\author{Wei-Tao Lu} \email{physlu@163.com}
\affiliation{School of Sciences, Nantong University, Nantong 226019, China}
\author{Qing-Feng Sun} \email{sunqf@pku.edu.cn}
\affiliation{International Center for Quantum Materials, School of Physics, Peking University, Beijing, 100871, China}
\affiliation{Hefei National Laboratory, Hefei 230088, China}
\affiliation{Beijing Academy of Quantum Information Sciences, West Bld. $\#$3, No. 10 Xibeiwang East Road, Haidian District, Beijing 100193, China}

\begin{abstract}
Recently, antiferromagnetic (AF) materials have attracted rapid attention, because they are considered as outstanding candidates to replace the widely used ferromagnets in the next generation of spintronics.
We propose a magnetoresistance model based on the quantum anomalous Hall effect in an AF system, which is protected by the topological Chern number.
By regulating the AF exchange field and an electric field, the system can be controlled between the quantum spin Hall insulator (QSHI) phases and the quantum anomalous Hall insulator (QAHI) phases. As a result, a QAHI/QSHI/QAHI junction can be formed.
In the QAHI region, the spin orientation of chiral edge state can be manipulated by tuning the AF exchange field and the electric field.
Therefore, the spin directions of two QAHIs in the junction can have parallel and antiparallel configurations.
The conductances of two configurations offered by chiral edge states are significantly different, and this is a magnetoresistance effect that can be electrically controlled.
Because of the topological invariance, the magnetoresistance plateaus are robust to the size effect and the disorder.
\end{abstract}
\maketitle

\section{Introduction}

The quantum anomalous Hall effect (QAHE) refers to the quantized Hall effect realized in a system without an external magnetic field, which is a topologically nontrivial phase characterized by a finite Chern number \cite{Xue1,CZChang}. Different from the helical edge states of quantum spin Hall effect (QSHE) \cite{Zhang3,Kane,Sun,addref5}, the edge states of QAHE are chiral.
The first model for QAHE was proposed by Haldane on honeycomb lattice \cite{Haldane}. Subsequently, the QAHE was predicted to occur in several candidate systems, including mercury-based quantum wells \cite{Zhang2}, magnetic topological insulators \cite{Fang, Xue2, Jiang, YDeng}, graphene \cite{Yao1,Yao2,PHogl,JLiu}, silicene \cite{Ezawa,Ezawa2,Yao3}, and other honeycomb materials \cite{Yan,Zhou,Nikolic,Xiang,LiuF2,Lu3}.
In silicene with an intrinsic spin-orbit coupling, the spin-polarized QAHE \cite{Ezawa2} and valley-polarized QAHE \cite{Yao3} could be realized when the time-reversal symmetry is broken by the exchange field. In recent experiments, the QAHE was first observed at extremely low temperatures in a thin film of chromium-doped (Bi,Sb)$_2$Te$_3$, a magnetic topological insulator \cite{Xue2}.
The chiral edge states carry dissipationless electric current owing to robustness against backscattering, and so QAHE has great potential applications in low-power consumption electronics. It has been demonstrated that the spin filters \cite{Ezawa3,Yao4,Xie} and field-effect transistors \cite{Ezawa4, Liu} could be generated by manipulating the topological edge states in many two-dimensional topological insulators.

On the other hand, magnetoresistance (MR) effect is a key concept in spintronics, which has applications in digital storage and magnetic sensor technologies.
Recently, a number of investigations on MR effect in two-dimensional materials with various ferromagnetic contacts have been reported \cite{Kim,Rossier,Honda,ZhangY,LiuF,Saxena,Lu,Chshiev,Lu2,LuJ, Yokoyama}.
Due to the unique symmetry of the band structure, the graphene nanoribbon device exhibits large MR values controlled by ferromagnetic electrodes \cite{Kim}. Based on the antiferromagnetic (AF) ground state in a short zigzag ribbon, the conductance could change dramatically as the magnetic orientation at the ribbon edges goes from parallel to antiparallel configurations by the magnetic field, leading to a giant MR \cite{Rossier, LuJ}.
The formation of spin-polarized quantum well states could give rise to a negative MR in the ferromagnetic MoS$_2$ junction \cite{Lu}. An anomalous MR occurs in a ferromagnet/ferromagnet junction on topological insulator, where the conductance depends strongly on the in-plane direction of the magnetization \cite{Yokoyama}.
At the current stage, the MR device is mainly controlled by the ferromagnets.

Compared with ferromagnets, antiferromagnets represent a more common form of magnetically ordered materials. Recent experiments demonstrate that an AF memory can be both written and read electrically \cite{Wadley}.
AF materials have great potentials in reducing the device size and power consumption, and thus become outstanding candidates for the next generation of spintronic applications \cite{Baltz, Smejkal2,addref1,addref2}.
Recently, QAHE has been observed in the intrinsic AF topological insulator MnBi$_2$Te$_4$ which is a layered ternary tetradymite compound that consists of Te-Bi-Te-Mn-Te-Bi-Te septuple layers \cite{YDeng}.
The QAHE could also be realized by coupling graphene to an AF insulator that provides both spin-orbit coupling and broken time-reversal symmetry \cite{Qiao, PHogl}.
AF spintronics initially focused on AF analogues of ferromagnetic spin valves and tunneling junctions.
Many works on the spin-transfer torque and MR effect have been reported in the AF tunnel junctions which are composed of two AF layers separated by a nonmagnetic barrier layer \cite{Nunez, Nunez2, YXu, Merodio, Saidaoui, Smejkal, LWang, DFShao, JDong, Stamenova}.
Importantly, it was found that the spin-transfer torque and the magnetization orientation of the AF materials can be manipulated by an electric bias \cite{Baltz, Smejkal2}.
N\'{u}\~{n}ez et al. predicted theoretically that a giant MR could occur in the AF tunnel junction \cite{Nunez}, which has been demonstrated in experiments \cite{LWang}.
By switching the N\'{e}el vector, a giant MR effect of $~500\%$ can be achieved in the RuO$_2$/TiO$_2$/RuO$_2$ AF tunnel junction using the AF metal RuO$_2$ as electrodes \cite{DFShao}.
In the CuMnAs/GaP/CuMnAs AF tunnel junctions, different magnetization states of the junction can be read by standard MR, while the MR depends on
the microscopic details of the interface \cite{Stamenova}.
Generally, the MR effect is sensitive to the finite size effect and the disorder.

In this work, we aim to study a topologically protected MR effect in the AF junction.
Compared with the previously studied conventional MR effect, the MR effect in this work arises from a spin-polarized QAHE and it is protected by topological invariance.
By regulating the electric field, the spin-polarized QAHE can be realized and its spin-polarization orientation can be switched.
Therefore, a junction composed of two QAHE regions can be manipulated, and their spin can be adjusted between parallel and antiparallel configurations.
From the different conductances of these two configurations, an MR effect is reflected.
The MR plateau in the QAHE regime is very remarkable and stable, robust to the size effect and disorder.
Furthermore, the generated MR effect can be easily controlled by electric field, which is advantageous for designing low-power consumption spintronic devices.

The paper is organized as follows. In Sec. II, we propose the tight-binding model and study the Chern number for quantum spin Hall insulator (QSHI) under the AF exchange field and electric field. In Sec. III, we present numerical results of the MR model which is composed of quantum anomalous Hall insulator (QAHI) and QSHI. Finally, a conclusion is given in Sec. IV.

\section{Theoretical Formulation}

We study the QSHI in a generic buckled honeycomb lattice (such as silicene and germanene nanoribbons) which could exhibit various topological phases such as QSHE and QAHE \cite{Ezawa,Ezawa2,Yao3}.
In the tight-binding representation, the Hamiltonian of the buckled honeycomb system under the AF exchange field $M_{AF}$ and the electric field $E_z$ can be written as:
\begin{eqnarray}
  H &=& -t \sum_{\langle i,j \rangle, \alpha} c_{i \alpha}^{\dag} c_{j \alpha} + i \frac{\lambda}{3\sqrt{3}}
    \sum_{\langle\langle i,j \rangle\rangle, \alpha, \beta} v_{ij} c_{i \alpha}^{\dag} (\sigma_z)_{\alpha\beta} c_{j \beta} \nonumber\\
&& + M_{AF} \sum_{i,\alpha} \xi_i c_{i \alpha}^{\dag} (\sigma_z)_{\alpha\alpha} c_{i \alpha} - l E_z \sum_{i,\alpha} \xi_i c_{i \alpha}^{\dag} c_{i \alpha}.
\end{eqnarray}
The first term describes the nearest-neighbor hopping with the hopping energy $t$, $c_{i \alpha}^{\dag}$ ($c_{i \alpha}$) is the electronic creation (annihilation) operator with spin $\alpha$ at site $i$, and $\langle i,j \rangle$ denotes the sum over the nearest-neighbor sites. The second term is intrinsic spin-orbit coupling with strength $\lambda$ which involves spin dependent next nearest-neighbour hopping, $\langle\langle i,j \rangle\rangle$ denotes the sum over the next nearest-neighbor sites, $\sigma_z$ is the Pauli matrix associated with spin degree of freedom, and $v_{ij} = +1 (-1)$ if the next nearest-neighboring hopping is anticlockwise (clockwise) with respect to the positive $z$ axis. The third term represents the AF exchange field $M_{AF}$ with the magnetization perpendicular to the plane and $\xi_i=\pm 1$ for $i=A, B$ sites. The last term is the staggered sublattice potential $l E_z$ arising from the electric field $E_z$ perpendicular to the plane and $l$ is the buckling height. The constant $l=0.035e{\AA}$, $0.046e{\AA}$, and $0.055e{\AA}$ in silicene, germanene, and stanene, respectively \cite{Tsai}. The electronic property of buckled honeycomb can be naturally tuned by the electric field.

The AF exchange field is a staggered exchange field with opposite sign on $A$ and $B$ sublattices. The AF order could be realized via proximity effect by depositing the honeycomb lattice to an antiferromagnet \cite{Qiao, PHogl}. Alternatively, it can be also realized by sandwiching the honeycomb lattice between two different ferromagnets due to the buckled structure \cite{Ezawa2}. The magnetization orientation of the AF order can be controlled by the electric bias \cite{Baltz, Smejkal2} and by the two ferromagnets \cite{Ezawa2}.
Thus, based on the proposed model, it is possible to realize an AF MR device.

The two-terminal conductance $G$ for an electron with energy $E$ through the device can be calculated by the nonequilibrium Green's function method and the Landauer-B\"{u}ttiker formula as \cite{addref3,addref4}
\begin{eqnarray}
G(E) = \frac{e^2}{h} Tr [\Gamma_L(E) G^r(E) \Gamma_R(E) G^a(E)],
\end{eqnarray}
where $\Gamma_{L,R}(E) = i [\Sigma_{L,R}(E) - \Sigma^{\dag}_{L,R}(E)]$ is the linewidth function and $G^r(E) = [G^a(E)]^{\dag} = 1 / (E - H_c - \Sigma_L - \Sigma_R)$ is the retarded Green's function with the Hamiltonian in the scattering region $H_c$. $\Sigma_{L,R}$ is the self energy caused by the coupling between the center and lead regions.

The low-energy effective Hamiltonian can be derived from the tight-binding Hamiltonian. It is described by the Dirac theory around the Dirac points:
\begin{eqnarray}
H = \hbar v_F (\eta \tau_x k_x + \tau_y k_y) + (\eta s \lambda + s M_{AF} - l E_z) \tau_z,
\end{eqnarray}
where $\eta = \pm 1$ denotes the $K$ and $K'$ valleys, $s = \pm 1$ denotes spin-up and spin-down states, and $\tau_{x,y,z}$ are Pauli matrices
denoting sublattice space.
Then the band structure for bulk materials can be obtained as
\begin{eqnarray}
E(k) = \pm \sqrt{(\hbar v_F)^2 (k_x^2 + k_y^2) + (\eta s \lambda + s M_{AF} - l E_z)^2}.
\end{eqnarray}
The Chern number $C$ determines the topological properties of the phase, which can be obtained by integrating the Berry curvatures over the first Brillouin zone \cite{Xiao}:
\begin{eqnarray}
C = \sum_{\eta, s} C_{\eta s} = \sum_{\eta, s} \frac{\eta}{2} sgn (\eta s \lambda + s M_{AF} - l E_z),
\end{eqnarray}
and the quantized Hall conductance is characterized by $\sigma_{xy} = C e^2 / h$.

\begin{figure}
\includegraphics[width=8.0cm,height=4cm]{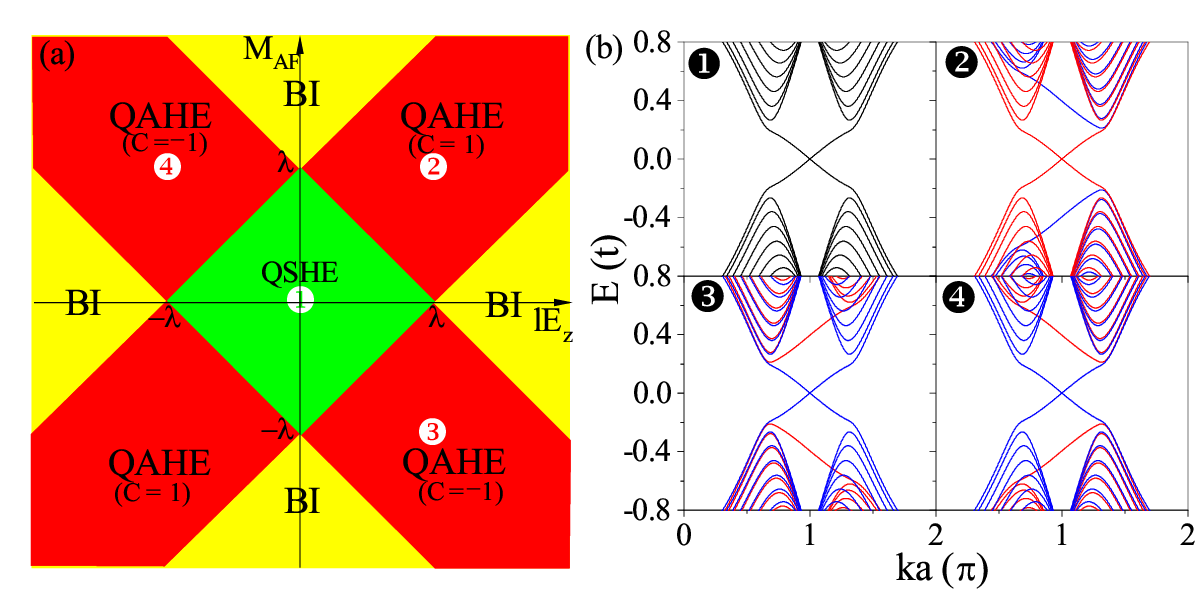}
\caption{ (a) Phase diagram of the QSHI in the ($l E_z, M_{AF}$) space. (b) Band structure of the system labeled by white points in (a) with $\lambda=0.2t$, where the red, blue, and black curves are for spin-up, spin-down, and spin-degenerate electrons, respectively.}
\end{figure}

Based on equations (1) and (5), we analyze the phase diagram
of QSHI in the presence of AF exchange field $M_{AF}$
and electric field $E_z$, as shown in Fig. 1(a).
For $|l E_z|+|M_{AF}|<\lambda$ (green area), the system is the QSHI with Chern number $C=0$ and
$C_{K\uparrow}+C_{K'\uparrow} = -C_{K\downarrow}-C_{K'\downarrow} =1$.
For $||l E_z|-|M_{AF}||>\lambda$ (yellow areas), the system is the band insulator (BI) with $C=0$ and $C_{K\uparrow}+C_{K'\uparrow} = C_{K\downarrow}+C_{K'\downarrow} =0$.
When $||l E_z|-|M_{AF}||<\lambda<|l E_z|+|M_{AF}|$ (red areas), the system undergoes a phase transition from QSHI phase with $C=0$ to QAHI phase with $C=\pm 1$.
In particular, the system has spin-up polarized QAHE in the first and third quadrants with $C=1$, while it is spin-down polarized QAHE in the second and fourth quadrants with $C=-1$. The most exciting finding is that the spin-polarization orientation of QAHE could be switched by tuning the sign of $M_{AF}$ or $E_z$. In order to clearly see the effect of $M_{AF}$ and $E_z$ on the QAHE, Fig. 1(b) displays four band structures labeled by four white points in the phase diagram of Fig. 1(a) where $l E_z=M_{AF}=0$, $l E_z=M_{AF}=\lambda$, $l E_z=-M_{AF}=\lambda$, and $-l E_z=M_{AF}=\lambda$, respectively. In the absence of external field ($l E_z=M_{AF}=0$) in Fig. 1(b1), the system is a QSHI associating with the helical edge states at the boundaries and the spin is degenerate. When $l E_z=M_{AF}=\lambda$ in Fig. 1(b2), the band gap for spin down is opened up and its edge state is destroyed. Thus, the system has spin-up polarized QAHE with the chiral edge state devoted by spin-up electron. The band gap is $2(|l E_z|+|M_{AF}|-\lambda)$ which can be controlled by $E_z$ and $M_{AF}$. Interestingly, the system would become spin-down polarized QAHI when we only reverse the exchange field $M_{AF}$ (or the electric field $E_z$), as shown in Fig. 1(b3) [or Fig. 1(b4)], implying that the spin index of the edge states has been switched and the Chern number changes from $C=1$ to $C=-1$. Such an important character can be used to design an MR device with topological invariance.

\section{Magnetoresistance Device and Results}

\begin{figure}
\includegraphics[width=7.0cm,height=5.0cm]{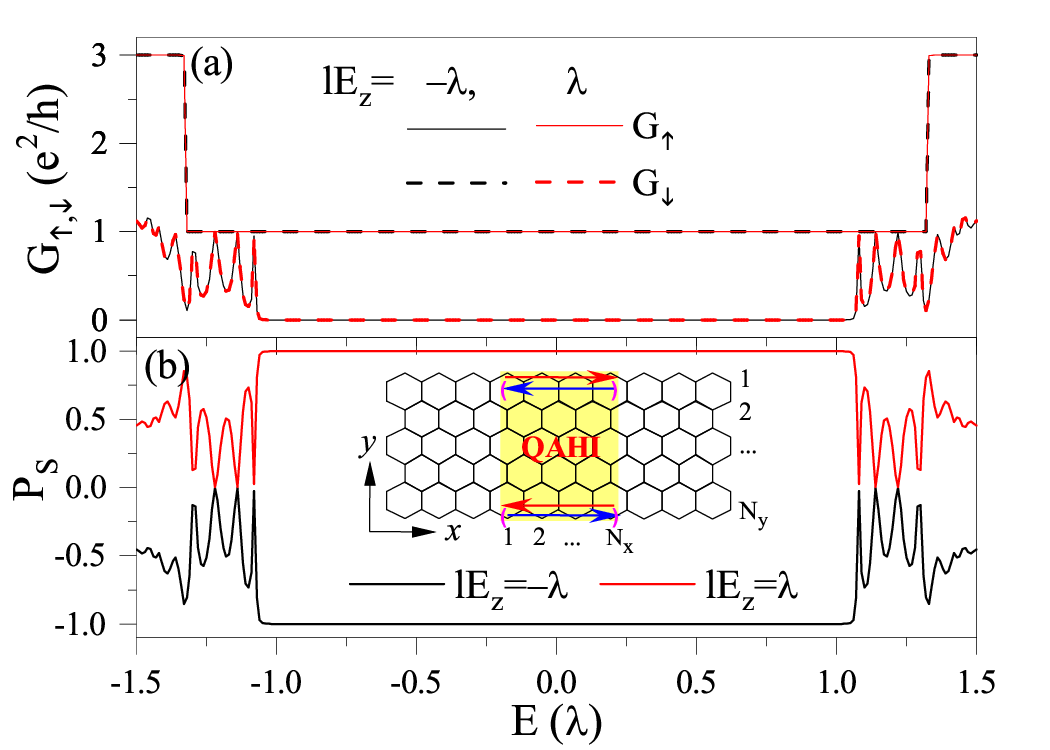}
\caption{ (a) Spin-dependent conductances $G_{\uparrow, \downarrow}$ and (b) spin polarization $P_S$ versus Fermi energy $E$ through a spin filter by virtue of spin-polarized QAHE.
The inset in (b) is the schematic diagram of spin filter. $M_{AF}$ and $E_z$ are only applied to the central yellow area. $N_x$ is the number of unit cells for the QAHI in the $x$ direction, and $N_y$ describes the width of the ribbon in the $y$ direction.
The red (blue) arrows are edge states for spin-up (spin-down) electron.
Here, $M_{AF}=\lambda$, $N_y=24$, and $N_x=50$.}
\end{figure}

Before discussing the MR effect, we present the spin polarization
in a spin filter [see the inset of Fig. 2(b)]
constructed by the spin-polarized QAHI, as shown in Fig. 2.
The spin polarization $P_S$ is defined as $P_S=(G_{\uparrow}-G_{\downarrow})/(G_{\uparrow}+G_{\downarrow})$.
Observably, both conductances $G_{\uparrow, \downarrow}$ and their spin polarization $P_S$ have the platforms in the region $|E|<\lambda$.
When $l E_z=M_{AF}=\lambda$ in the region $|E|<\lambda$,
the conductance $G_{\uparrow}$ is $e^2/h$ devoted by the spin-up polarized edge state of the QAHI in the central yellow area [see the inset of Fig. 2(b)],
while $G_{\downarrow}$ is zero due to the band gap of spin down,
leading to $P_S=1$. The irregular oscillation of $G_{\downarrow}$ in the region $|E|>\lambda$ originates from the asymmetric band structure [see Fig. 1(b2)].
The platforms with $G_{\downarrow}=e^2/h$, $G_{\uparrow}=0$, and $P_S=-1$ can be achieved by the spin-down polarized edge state when the electric field or the AF exchange field is reversed.

\begin{figure}
\includegraphics[width=7.0cm,height=7.0cm]{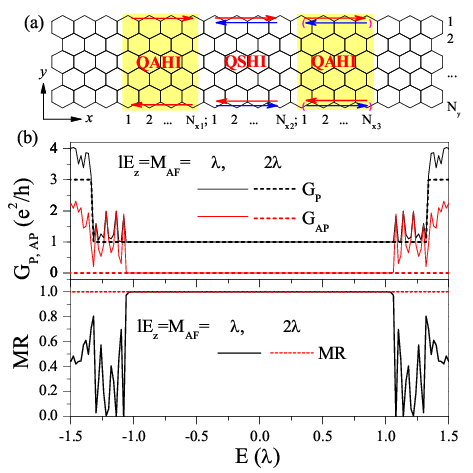}
\caption{ (a) Schematic diagram of the MR model composed of QSHI
(white areas) and QAHI (yellow areas).
(b) Conductances $G_{P, AP}$ and MR versus Fermi energy $E$ in the model.}
\end{figure}

Based on the above analysis, we propose an MR model, i.e., the QAHI/QSHI/QAHI junction, as shown in Fig. 3(a), which is manipulated by the AF exchange fields $M_{AF}$ and the electric fields $E_z$. Importantly, the junction is composed of QSHI and spin-polarized QAHI, and so it is topologically protected.
The fields $M_{AF}$ and $E_z$ are applied to the two yellow areas of the model, while they are absent in other areas.
The directions of $M_{AF}$ and $E_z$ on the right can be regulated.
If the AF exchange fields on the left and the right are in the same (or opposite) direction, the structure would show the parallel (or antiparallel) configuration, implying that the spin polarizations of edge states on the left and the right are the same (or opposite). In the following calculation, we set $M_{AF}^L=M_{AF}^R=M_{AF}$ in the parallel case, while $M_{AF}^L=-M_{AF}^R=M_{AF}$ in the antiparallel case. The parameter values are set as $N_y=24$ and $N_{x1}=N_{x2}=N_{x3}=50$ for convenience, unless otherwise noted.

The conductance of the parallel and antiparallel cases can be written as $G_{P,AP}=G_{P,AP}^{\uparrow} + G_{P,AP}^{\downarrow}$. The MR ratio is defined as: $MR=(G_P-G_{AP})/G_P$.
Fig. 3 (b) presents the conductances $G_{P, AP}$ and the MR as a function of Fermi energy $E$ for different values of $M_{AF}$ and $l E_z$.
In the parallel case, both QAHIs on the left and the right are spin-up polarized and their Chern numbers are the same with $C=1$.
As a consequence, the conductance is $G_P=e^2/h$ contributed by the edge states of spin-up electron in the region $|E|<\lambda$,
corresponding to the QAHE in the energy gap for bulk materials.
Oppositely, in the antiparallel case, due to the reversed spin-polarized QAHE and the different Chern numbers between the left and the right,
the transport of the edge states is forbidden and so the conductance $G_{AP}=0$. Thus, a remarkable MR effect is realized in the QAHE regime which is protected by the topological Chern number. Out of the region $|E|<\lambda$, more and more bulk states would devote to the current.
The difference between $G_P$ and $G_{AP}$ becomes unnoteworthy,
and so the MR trends toward zero. With the increase of $M_{AF}$ and $E_z$, the MR plateau is broadened [see Fig. 3(b)], because the energy gap of one spin with the gapped state in QAHE is enlarged.

\begin{figure}
\includegraphics[width=8.0cm,height=8.0cm]{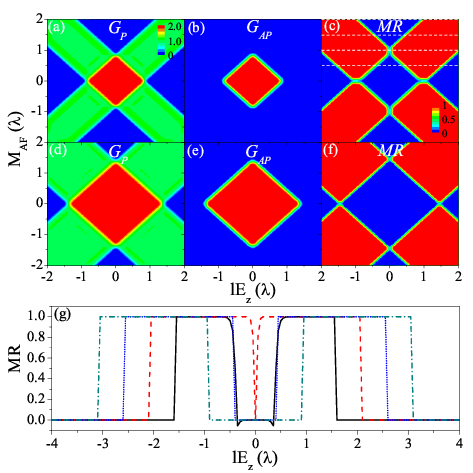}
\caption{ Conductances $G_{P, AP}$ and the MR in the ($l E_z, M_{AF}$) plane at (a-c) $E=0.005\lambda$ and (d-f) $E=0.5\lambda$. The unit of conductance is $e^2/h$. (g) MR versus the staggered sublattice potential $l E_z$ labeled by the white dashed lines in (c). The solid, dashed, dotted, and dash-dotted curves are for $M_{AF}=0.5\lambda$, $1.0\lambda$, $1.5\lambda$, and $2.0\lambda$, respectively.}
\end{figure}

For purpose of an overview on the effect of $E_z$ and $M_{AF}$,
Fig. 4 shows the contour plot of conductances $G_P$, $G_{AP}$ and MR in the ($l E_z, M_{AF}$) plane for different values of Fermi energy $E$.
One can clearly see that the contour plots of conductance and MR are accurately consistent with the phase diagram in Fig. 1(a).
In the QSHE regime, the conductance for both $G_P$ and $G_{AP}$ is $2e^2/h$ due to the helical edge states, and so the MR is zero [see Figs. 4(a)-4(c)].
In the BI regime, there is no edge states, so $G_P=G_{AP}=0$, and $MR=0$.
In the QAHE regime, $G_{AP}$ is zero, while $G_P$ is $e^2/h$ due to the chiral edge state devoted by the spin-up or spin-down electron.
Therefore, a topologically protected MR is achieved in the QAHE regime
due to the spin-polarized QAHE. Obviously, the conductance $G_P$ is spin polarized in the QAHE regime.
For a large Fermi energy $E$, such as $E=0.5\lambda$ in Figs. 4(d)-4(f), the topologically protected MR and spin polarized current could also be realized in the QAHE regime.
Because of the insulating states, the conductance region in the QSHE/QAHE regimes and the MR region in the QAHE regime are broadened, the size of which is increased by $0.5\lambda$.
In order to see the MR clearly, Fig. 4(g) exhibits MR as a function of $l E_z$ with $M_{AF}=0.5\lambda$, $1.0\lambda$, $1.5\lambda$, and $2.0\lambda$, as labeled by the white dashed lines in Fig. 4(c). Dramatically, an MR plateau with $MR=1.0$ is realized in certain regions of $l E_z$, otherwise, $MR=0.0$, which can be controlled by $M_{AF}$.
The width of the MR plateau is about $2\lambda$ when $|M_{AF}| \geq \lambda$.
This means that MR can be adjusted by the staggered sublattice potential $l E_z$ which can easily be controlled by a gate voltage in the experiment.
Furthermore, it is interesting to find that the small negative MR appears near the boundary of QAHE and QSHE regimes [see the solid curve in Fig. 4(g)], due to the close of the bulk gap at the QAHE-QSHE boundary.
The bulk states would contribute to the conductance and give rise to the small negative MR. This small negative MR is sensitive to the Fermi energy and the structural parameters, and it is not robust against the disorder.
These properties suggest that the QAHI/QSHI/QAHI junction could work as an electrically controlled MR switch.

\begin{figure}
\includegraphics[width=8.0cm,height=5.0cm]{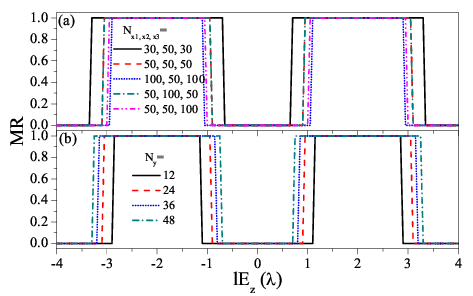}
\caption{ MR versus staggered sublattice potential $l E_z$ (a) for different values of the widths $N_{x1,x2,x3}$ with $N_y=24$ and (b) for different values of the ribbon width $N_y$ with $N_{x1}=N_{x2}=N_{x3}=50$. Here, $E=0.005\lambda$ and $M_{AF}=2\lambda$.}
\end{figure}

The finite size effect is crucial for the device application. It's worth noting that because the MR effect is protected by the topological Chern number, it is insensitive to the finite size effect of the device. The widths $N_y$, $N_{x1}$, $N_{x2}$, and $N_{x3}$ [see Fig. 3(a)]
have no distinct effect on the conclusion.
Fig. 5 discusses the dependence of MR on $l E_z$ for different values of (a) the widths $N_{x1,x2,x3}$ and (b) the width $N_y$ when $E=0.005\lambda$ and $M_{AF}=2\lambda$.
The position of MR plateau is approximately in the region $M_{AF}-\lambda < |l E_z| < M_{AF}+\lambda$, i.e., $\lambda < |l E_z| < 3 \lambda$, corresponding to the QAHE regime. From the solid curve, dashed curve, and dotted curve in Fig. 5(a) one can see that as the barrier widths $N_{x1,x3}$ decrease, the MR plateau is widened a little due to the tunneling effect of wavefunction. The well width $N_{x2}$ has no effect on MR [comparing the dashed curve with dash-dotted curve in Fig. 5(a)]. The asymmetry of the device [i.e., $N_{x1} \neq N_{x3}$, see the dash-dot-dotted curve in Fig. 5(a)] cannot destroy the MR plateau as well. It is well known that the number of bands is $2 N_y$ in the nanoribbon.
As $N_y$ increases, more bands appear in the considered energy region and the band of edge state is shifted down indistinctively. Consequently, the critical value of $l E_z$ for conductance in QAHE regime is modulated and the MR plateau could be broadened slightly with the increase of $N_y$, as shown in Fig. 5(b). Distinctly, the MR effect is stable and insensitive to the structural parameters of the system.

\begin{figure}
\includegraphics[width=8.0cm,height=6.0cm]{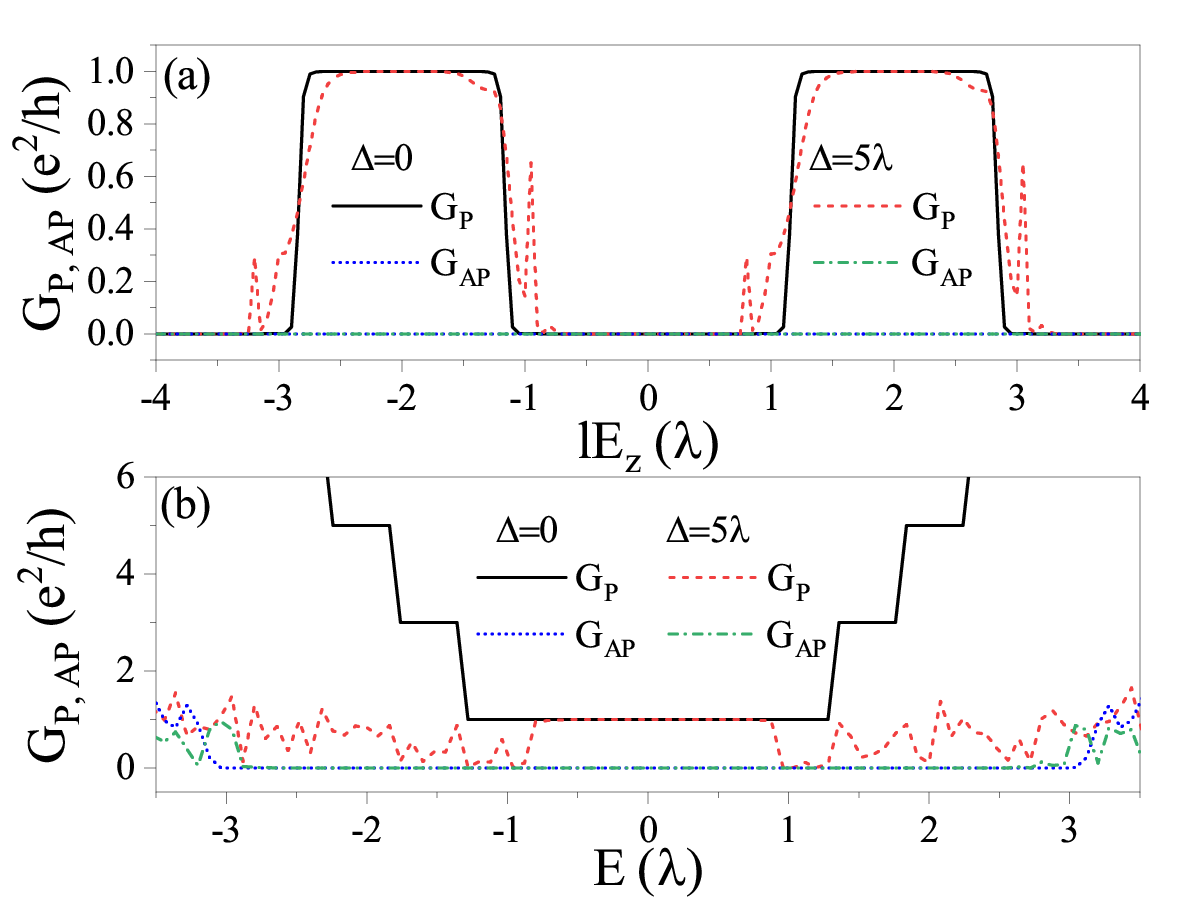}
\caption{ (a) Conductances $G_{P, AP}$ versus staggered sublattice potential $l E_z$ at $M_{AF}=2\lambda$ and $E=0.005\lambda$. (b) Conductances $G_{P, AP}$ versus Fermi energy $E$ at $l E_z=M_{AF}=2\lambda$.
The size of the center region is $N_y=24$ and $N_{x1}=N_{x2}=N_{x3}=50$.}
\end{figure}

\begin{figure}
\includegraphics[width=8.0cm,height=8.0cm]{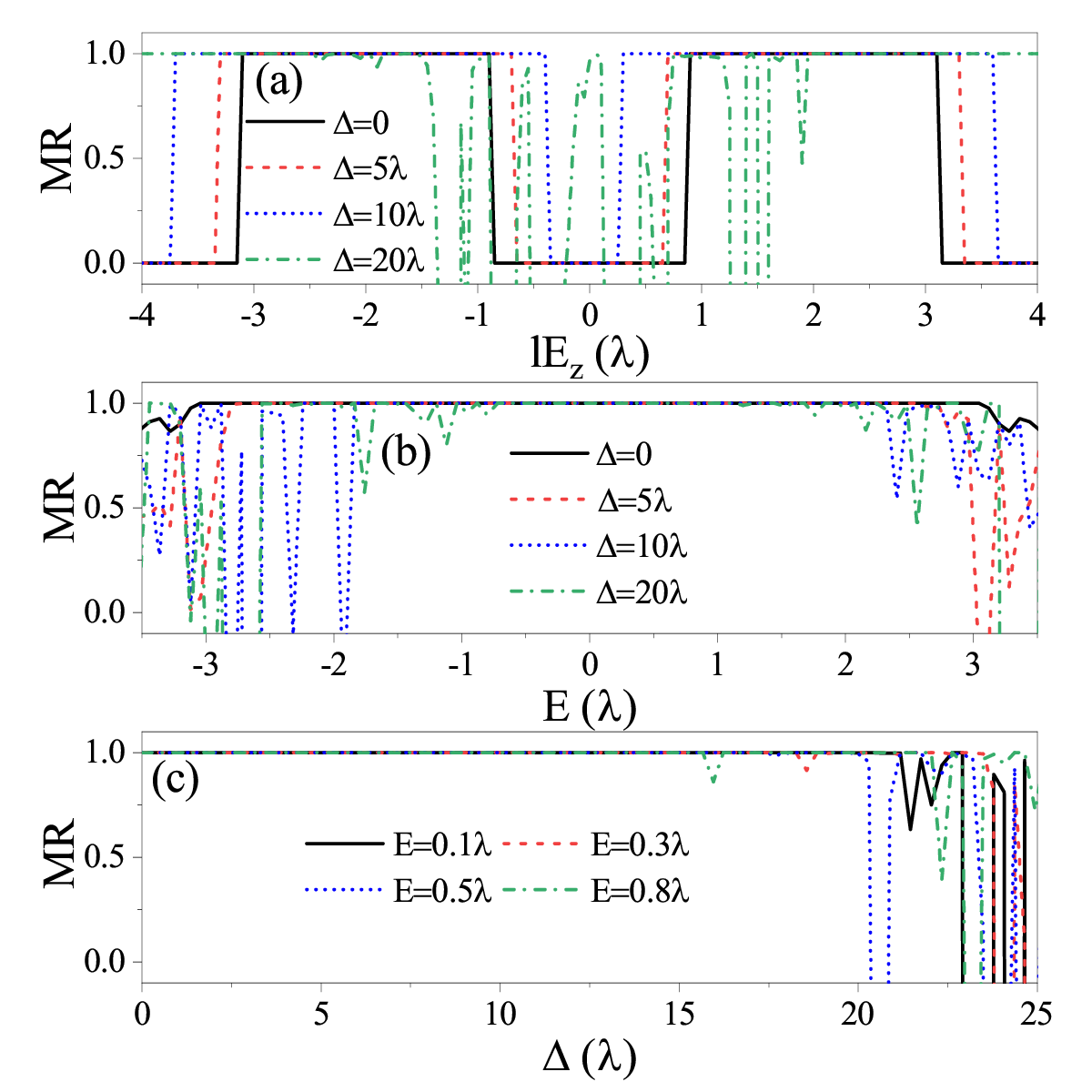}
\caption{ (a) MR versus staggered sublattice potential $l E_z$ at $M_{AF}=2\lambda$ and $E=0.005\lambda$. MR versus (b) Fermi energy $E$ and (c) disorder strength $\Delta$ at $l E_z=M_{AF}=2\lambda$.}
\end{figure}

We also examine the effect of disorder on the topologically protected MR.
In the experiment, disorder and impurity exist inevitably.
We assume that the on-site disorder potential exists in the central scattering region, which is randomly distributed in the range [$-\Delta/2, \Delta/2$] with the disorder strength $\Delta$.
The center region has a width of $N_y=24$ atomic sites and a length of $N_{x1}+N_{x2}+N_{x3}=150$ atomic sites.
The conductance and the MR are averaged over up to $3600$ random disorder configurations.
Figs. 6 and 7 show the conductances $G_{P, AP}$ and the MR in the presence of disorder, respectively.
From Fig. 6 one can see clearly that the quantum plateaus of $G_P=e^2/h$ contributed by the edge states are robust against the disorder due to the topological invariance [see the curves in the region $\lambda<|l E_z|<3\lambda$ of Fig. 6(a) and the curves in the region $|E|<\lambda$ of Fig. 6(b)].
However, the quantized conductance $G_P=ne^2/h$ with the integer $n>1$ offered by the bulk states is destroyed severely by the disorder, as shown in Fig. 6(b).
On the other hand, $G_{AP}$ still maintains its zero value in the band gap region when the disorder appears.

The MR as a function of the staggered sublattice potential $l E_z$ for different values of disorder strength $\Delta$ is shown in Fig. 7(a).
When the disorder is present, one can notice that the MR plateau still
holds primely. The weak disorder has no effect to the MR.
Most dramatically, the MR plateau is widened as $\Delta$ increases.
The reason behind is that the conductance region of $G_P$ is broadened by the disorder while $G_{AP}$ is always zero [see Fig. 6(a)].
However, with the further increase of $\Delta$,
such as $\Delta=20\lambda$ which is much larger than the energy gap
of the bulk materials, the MR plateau will be destroyed.
Meanwhile, a negative MR is induced by the disorder.
In the energy space, the MR plateau occurs in the region $|E| \leq \lambda$, i.e., in the QAHE regime.
As shown in Fig. 7(b), the plateau is well-preserved for a weak disorder $\Delta < 10 \lambda$. For a strong disorder,
the plateau will be shortened due to the effect of bulk states.
However, the plateau structure of MR is still perfect near the zero energy,
for example, in the region $|E| \leq 0.7\lambda$ for $\Delta = 20 \lambda$ which is four times as large as the hopping energy $t$.
The continuous dependence of MR on the disorder strength $\Delta$
for different values of Fermi energy $E$ is displayed in Fig. 7(c).
It can be found that the quantum plateau maintains its quantized value
very well even when $\Delta$ reaches up to $15 \lambda$.
The results demonstrate that the MR effect is very robust against the disorder because of the topological invariance and the chiral edge states of the system.

\begin{figure}
\includegraphics[width=8.0cm,height=4.0cm]{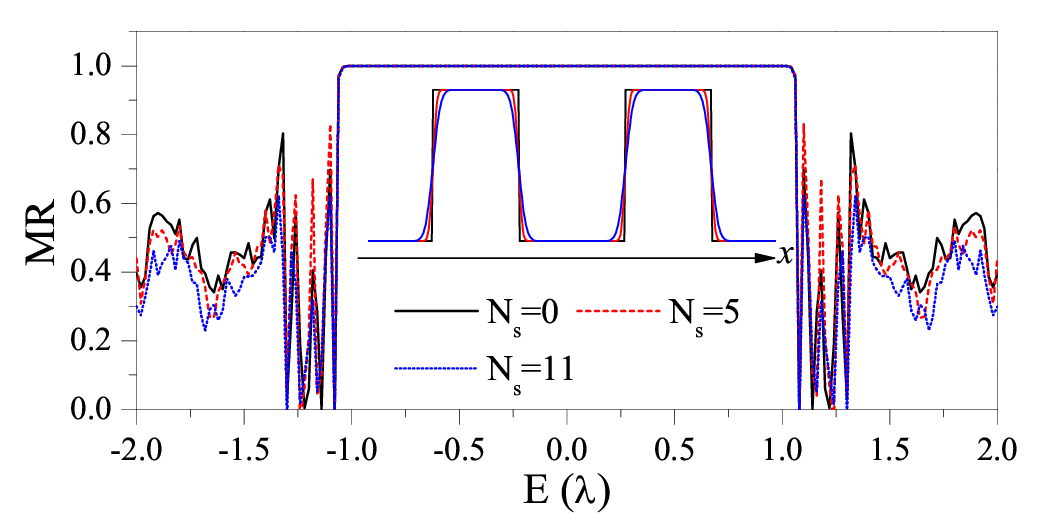}
\caption{ MR versus Fermi energy $E$ for different values of $N_s$.
The schematic diagram of the parameters $M_{AF}$ and $E_z$ for the smooth QAHI/QSHI/QAHI junction is shown in the inset.
Here, $M_{AF}=lE_z=1.0\lambda$, $N_y=24$, and $N_{x1}=N_{x2}=N_{x3}=50$.}
\end{figure}

In the above investigation, an ideal model with sharp boundary is employed to discuss the physics clearly.
However, in the experiment, the antiferromagnetic exchange field $M_{AF}$ and electric field $E_z$
would vary smoothly in space.
Below we consider a realistic QAHI/QSHI/QAHI junction and its profile of the fields $M_{AF}$ and $E_z$ is shown
in the inset of Fig. 8, which can be described by the error function.
There are four boundaries between QAHI and QSHI regions in the junction [see Fig. 3(a)].
For each smooth boundary, $N_s$ labels the number of unit cells where the fields $M_{AF}$ and $E_z$ vary smoothly along the $x$ direction.
$N_s=0$ corresponds to the sharp boundary in the ideal junction.
With the increase of $N_s$, the boundary becomes smoother and smoother.
Fig. 8 shows the MR as a function of Fermi energy $E$ for different values of $N_s$.
The result indicates that the smooth boundary has a certain influence on the MR contributed by the bulk states in the region $|E|>\lambda$.
However, in the region $|E|<\lambda$, no matter the boundary is smooth or sharp, the perfect MR platform contributed by the edge states always holds.
Prominently, the MR is robust to the smooth boundary.
The reason is that the smooth variations of the fields $M_{AF}$ and $E_z$ at the QAHI-QSHI boundary do not change the topological properties of the system, including QAHI and QSHI.

\section{Conclusion}
In summary, we theoretically propose a topologically protected MR effect in the AF system by using the tight-binding method. Unlike the MR effect studied so far, the origin of this MR is the spin-polarized QAHE.
Because the QAHE and Chern number can be regulated by the electric field, the MR properties of this device can be easily controlled.
The proposed MR effect is stable and robust against the disorder.
The research should be beneficial for designing spintronic devices based on topological states.

\section*{Acknowledgments}
We would like to thank Y. Mao for many helpful discussions.
This work was supported by the NSFC (Grants No. 11974153 and No. 11921005),
the Innovation Program for Quantum Science and Technology (2021ZD0302403), and the Strategic Priority Research Program of Chinese Academy of Sciences (Grant No. XDB28000000).

\end{CJK*}
\end{document}